# Low-fat diets and testosterone in men: systematic review and meta-analysis of intervention studies

Joseph Whittaker [a*] and Kexin Wu [b]

[a] The School of Allied Health and Community, University of Worcester, Henwick Grove, WR2 6AJ, United Kingdom. E-mail address: josephwhittaker1991@hotmail.co.uk

[b] Department of Statistics, University of Warwick, Coventry, CV4 7AL, United Kingdom.

*Corresponding author: Joseph Whittaker.

**Background:** Higher endogenous testosterone levels are associated with reduced chronic disease risk and mortality. Since the mid-20th century, there have been significant changes in dietary patterns, and men's testosterone levels have declined in western countries. Cross-sectional studies show inconsistent associations between fat intake and testosterone in men.

**Methods:** Studies eligible for inclusion were intervention studies, with minimal confounding variables, comparing the effect of low-fat vs high-fat diets on men's sex hormones. 9 databases were searched from their inception to October 2020, yielding 6 eligible studies, with a total of 206 participants. Random effects meta-analyses were performed using Cochrane's Review Manager software. Cochrane's risk of bias tool was used for quality assessment.

**Results:** There were significant decreases in sex hormones on low-fat vs high-fat diets. Standardised mean differences with 95% confidence intervals (CI) for outcomes were: total testosterone [-0.38 (95% CI -0.75 to -0.01) P = 0.04]; free testosterone [-0.37 (95% CI -0.63 to -0.11) P = 0.005]; urinary testosterone [-0.38 (CI 95% -0.66 to -0.09) P = 0.009]; and dihydrotestosterone [-0.3 (CI 95% -0.56 to -0.03) P = 0.03]. There were no significant differences for luteinising hormone or sex hormone binding globulin. Subgroup analysis for total testosterone, European and North American men, showed a stronger effect [-0.52 (95% CI -0.75 to -0.3) P < 0.001].

**Conclusions:** Low-fat diets appear to decrease testosterone levels in men, but further randomised controlled trials are needed to confirm this effect. Men with European ancestry may experience a greater decrease in testosterone, in response to a low-fat diet.

---

## 1. Introduction

Testosterone (T) plays a fundamental role in male physiology and reproductive health. Low endogenous T levels are associated with a higher risk of chronic disease [1–3]; and all-cause mortality [4]. Several studies in western and other modernised countries have found an age-independent secular decline in men's serum total testosterone (TT) of approximately 1% per year, beginning in the 1970s [5–12]. This downwards secular trend in T is only partly explained by the concurrent rise in BMI over the late 20th century [5,6]. The other factors responsible for this trend remain to be elucidated.

Western dietary patterns have changed substantially over the 20th century, with processed food consumption more than doubling in western countries [13,14]. The trend towards increased processed food and decreased whole food consumption, has contributed to a change in macronutrient intakes since the mid-20th century. In the USA, from 1965–71 fat intake decreased by 235kcal/day, and from 1965-91 carbohydrate intake increased by 263kcal/day [15]. These changes resulted in a 10.1% decrease in fat intake, as a percentage of total energy intake (TEI). From 1991-2011, macronutrient intakes remained relatively stable [15]. The decrease in fat intake was largely a response to dietary guidelines advising a reduction in fat intake, starting in the 1960s [15–17]. In the USA, the first evidence of a secular decline in T is from the 1980s [5], which is at the tail end of the changes in macronutrient intakes. This prompts the question of whether changes in fat intake may affect T levels.

Cross-sectional studies on the association between fat intake and T have produced conflicting results [18–22], possibly due to unmeasured confounding variables. However, the largest of these studies found that men adhering to a low-fat (LF) vs non-restricted diet had significantly lower serum TT (-32.7 ng/dL) [18]. General reviews on diet and sex hormones have only briefly covered the topic of dietary fat and T [23–26]. Moreover, these reviews have not included meta-analysis of homogeneous studies, have included studies with significant confounding variables, and missed eligible studies. Therefore, to investigate whether the association between LF diets and TT is causal, we conducted a systematic review and meta-analysis on intervention studies comparing the effect of a LF vs high-fat (HF) diet on men's TT and related hormonal markers.

---

*Abbreviations:* Chi$^2$, chi-squared; CI, confidence interval; DHT, dihydrotestosterone; FT, free testosterone; HF, high-fat; LF, low-fat; LH, luteinising hormone; MUFA, monounsaturated fatty acid; NA, North American; ND, no data; P:S, polyunsaturated to saturated fatty acid ratio; PUFA, polyunsaturated fatty acid; SA, South African; SD, standard deviation; SEM, standard error of the mean; SFA, saturated fatty acid; SHBG, sex hormone binding globulin; T, testosterone; TEI, total energy intake; TT, total testosterone; UT, urinary testosterone.

## 2. Methods

The review was structured using the PRISMA (preferred reporting items for systematic reviews and meta-analyses) 2009 checklist [27]. The review protocol was retrospectively registered with Research Registry, the unique identifying number is reviewregistry1085 [28]. The *Cochrane Handbook for Systematic Reviews of Interventions* was used as a guide throughout the review process [29].

### 2.1. Eligibility criteria

Studies were eligible if they met all of the following criteria:

1. Measurement of either TT, free testosterone (FT), urinary testosterone (UT), dihydrotestosterone (DHT), sex hormone binding globulin (SHBG), or luteinising hormone (LH).
2. Healthy adult male participants only (18+ years); to minimise participant variation in androgen metabolism due to disease or biological maturity.
3. Intervention at least 1 week, as the acute effects of food intake on androgen metabolism, may not reflect the long-term effects [30].
4. A difference in fat intake between diets ≥10% of TEI, to provide a meaningful distinction between diets on the basis of fat intake. Percentage rather than total fat intake was used to account for participant differences in energy intake requirements.
5. No factors which may alter androgen metabolism such as exogenous hormones, medications [31], dietary supplements [32,33], phytoestrogens [34], changes in exercise levels [35], protein >20% of TEI [36], carbohydrate <15% of TEI [37], weight loss >2kg [38], and a difference in TEI between diets ≥10% [38,39].

Studies which stated diets were isocaloric, but did not report precise TEIs were eligible. Hypogonadal participants were eligible, being a population of special interest, however no eligible studies contained them.

### 2.2. Search strategy

The following databases were searched from conception to 10th October 2020: MEDLINE, EMBASE, CINAHL, SPORTDiscus, Allied and Complementary Medicine Database, ClinicalTrials.gov, International Clinical Trials Registry Platform, Open Grey, and Google Scholar. Where possible the following search filters were applied: human, adult (18+ or 19+ years), male, intervention studies, and completed/ with results; no restrictions were put on language. A broad search was conducted on all databases except Google Scholar, using the search terms: (hypogonadism OR testosterone OR luteini* OR sex hormone* OR androgen*) AND (diet OR dietary OR macronutrient*). This was complimented by a precise search on Google Scholar limited to the first 200 results [40], using the search terms: men diet hypogonadism OR testosterone OR androgen OR hormones -mice -rats -cats -dogs -animals -mouse. Google Scholar was included as its addition to a MEDLINE and EMBASE search has been shown to increase overall recall by 3.1% [40]. However, since search filters were used a non-MEDLINE PubMed search could not be conducted, as non-MEDLINE results are not filterable [41].

The initial database search and screen was done by 1 author (J. W.), which yielded 2594 results (Fig. 1). 4 related reviews were found [23–26], and their references lists were searched manually. References lists of any eligible studies were also searched manually. 28 records were selected for full text review, 26 of which were derived from the initial search and 2 from reference lists. The first author (J. W.) assessed each full text article for eligibility and selected the final reports and studies for inclusion; the results of the full text screening were later reviewed by the second author near the end of the overall review process. 6 studies were selected for inclusion in the systematic review and meta-analysis, which were reported over 12 articles.

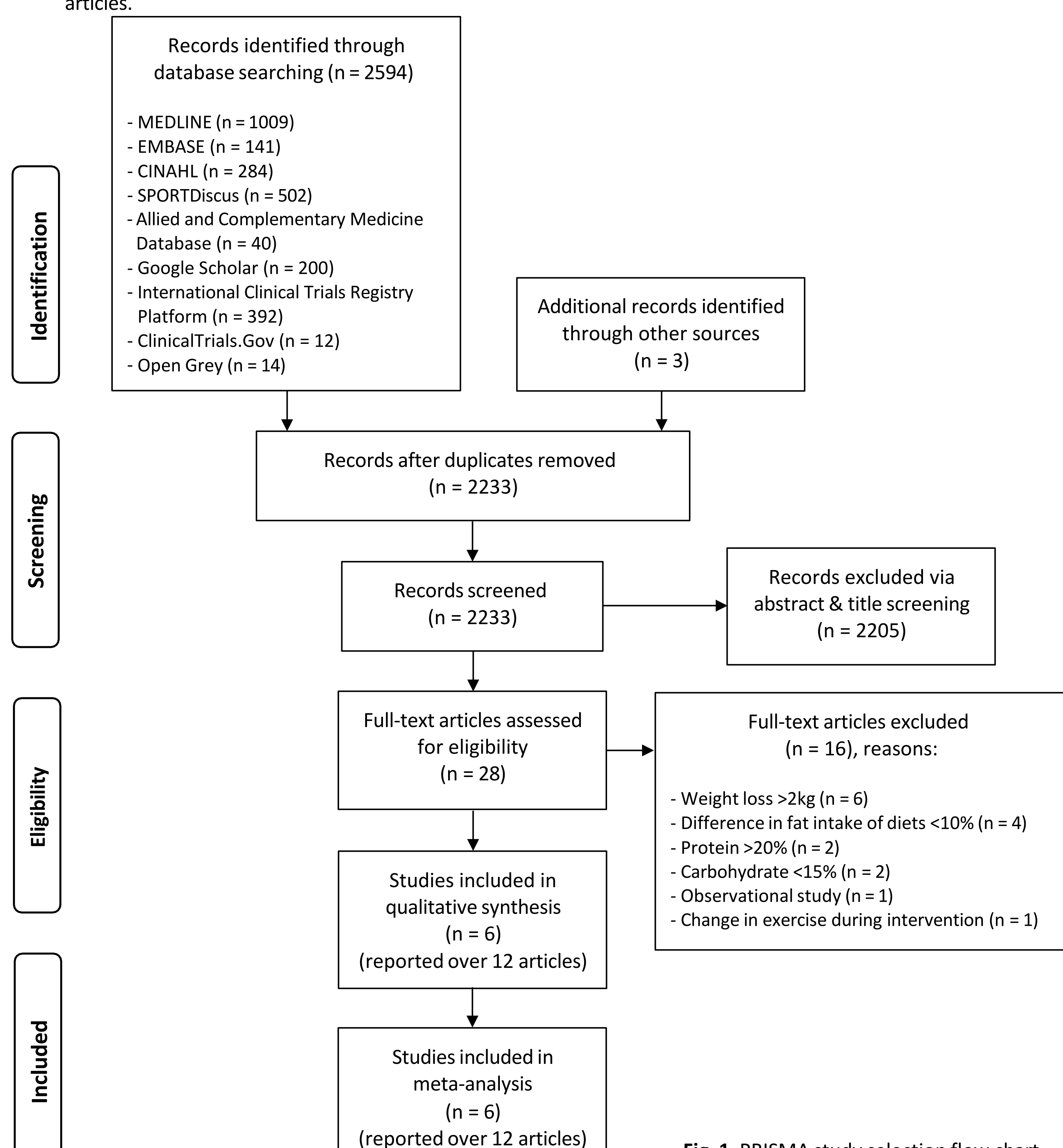

**Fig. 1.** PRISMA study selection flow chart.

### 2.3. Data extraction

1 author (J. W.) extracted the following data from each study: total sample size; sample size for each outcome; mean age; length of intervention; calorie and fat intake; bodyweight; outcomes for each diet with standard error of the mean (SEM), standard deviation (SD) or confidence intervals (CI); and P value for mean changes. The outcomes were TT, FT, UT, DHT, SHBG, and LH. Where data was presented only in graphs and the original authors could not provide it, the data was digitally extracted using the tool: Web Plot Digitizer [42]. All extracted data was triple checked against the original studies. Where possible and appropriate the studies' authors were contacted for additional information.

### 2.4. Data syntheses

When studies reported multiple outcome measurements from the same day, a mean and combined SD of these was used. Subgroups within studies were combined, and SEM and CIs were converted into SDs. To do this, formulas from the Cochrane handbook were used [43]. Hill 1980 featured 2 different interventions and samples respectively, North American (NA) hospital staff and rural South African (SA) farmers [44]. The study was split into 2 groups for analysis, as these 2 study arms could not be fairly combined. This meant there were 7 total samples for meta-analysis, from the 6 original studies. Additional details on how each study was handled can be found in Appendix A: Supplemental Methods.

### 2.5. Meta-analyses

All studies accounted for the diurnal rhythms of sex hormones, by ensuring the time of the blood sample was the same for each study condition. 2 studies used serum measurements [45,46], and 4 studies used plasma measurements [44,47–49], with different hormonal assays used across the studies. Therefore, to minimise differences due to blood samples and assays, standardised mean differences with 95% CIs were used for all outcomes. All of the studies were crossover trials, but did not show any significant period or carry-over effects. Therefore, the outcomes for HF and LF diets (mean ± SD) were used as the basis for the pooled effect estimates. The variance of within participant differences was not used for pooled effect estimates, as this would have required variance to be calculated based on assumptions for the majority of studies [50]. The meta-analyses were done using Cochrane's software: Review Manager [51]. DerSimonian and Laird's random effects model was used for all outcomes [52]. Chi-squared ($Chi^2$) and $I^2$ tests were used to measure heterogeneity, however since these tests were underpowered due to the low number of studies, forest plots were also visually inspected for heterogeneity. $I^2$ = 0 - 30% was not taken as conclusive evidence of low heterogeneity, but was interpreted alongside P values from $Chi^2$ tests and the visual inspection of forest plots. As a rule of thumb, P < 0.05 for Z tests were used as an indication of statistical significance, not an absolute measure of it, as suggested by Cochrane [53]. Standardised mean differences were interpreted as follows: 0.2 = small effect, 0.5 = moderate effect, and 0.8 = large effect [54].

### 2.6. Subgroup and sensitivity analyses

A post hoc subgroup analysis was done for TT western samples, as the only non-western sample [44], showed significant qualitative interaction. Western samples were defined as samples from North America or Europe. This yielded a reasonably ethnically homogeneous subgroup approximately 87.3% white/ European descent and 12.7% black NA. Another post hoc sensitivity analysis was conducted for TT, by taking out each study in turn and running the meta-analysis again. These results were informally compared to each other, as suggested by Cochrane [55].

An additional post hoc sensitivity analysis was done for TT, excluding the 2 most clinically diverse and highest risk of bias studies: Reed 1987 and Hill 1979 [47,49]. These studies had notably smaller sample sizes, and the LF diet in Reed 1987 had the lowest fat intake of all study diets by a margin of 7.1% of TEI [47] (Table 1). There was also limited data on participant age in Hill 1979 [49] (Table 1). Other post hoc subgroup analyses were conducted on the basis of ethnicity and study quality, but the results not presented due to potentially data dredging the small sample of studies.

### 2.7. Quality assessment

Since all of the studies were crossover trials, Cochrane's risk of bias crossover variant tool was used [56]. 5 of the 6 studies were not randomised [44–47,49]; therefore the randomisation bias questions were omitted for these studies. An additional bias domain was added to the Cochrane tool for the confounding variable of differences in micronutrient intake between HF and LF diets. Both authors conducted the risk of bias assessment, besides the custom bias domain which was done by the first author only (J. W.); a consensus for the final results was reached between authors. Afterwards, a risk of bias graph was generated using Cochrane's software: Risk of Bias Visualisation [57]. In this graph, studies were weighted using their weights from the TT meta-analysis. Forest plots were ordered from lowest to highest risk of bias.

**Table 1**
Characteristics of included studies[a]

| Study | Sample size | Age | HF diet Length (weeks) | LF diet Length (weeks) | HF diet Fat intake (% TEI) | LF diet Fat intake (% TEI) | LF vs HF diet Mean difference TEI (kcal/day) | Change in bodyweight (kg) | Risk of bias[b] |
|---|---|---|---|---|---|---|---|---|---|
| Dorgan 1996 [48,58] | 43 | 33.8 ± 1.4 | 10 | 10 | 40.7 ± 0 | 18.9 ± 0 | 24 (median) | -0.6 (median) | low |
| Wang 2005 [46] | 39 | 54.2 ± 0.5 | baseline diet | 8 | 37.9 ± 1 | 13.9 ± 0.3 | -30 | -1.1 | low |
| Hämäläinen 1984 [45,59,60] | 30 | 45.5 | 2 (baseline diet) | 6 | 40 | 25 | -179 | -1.1 | low |
| Hill 1980 NA [44,61–63] | 34 | 46.8 ± 1.1 | 2 | 3 | 40 | 25 | ND | -0.6 | medium |
| Hill 1980 SA [44,61–63] | 39 | 53.4 ± 2.3 | 3 | baseline diet | 40 | 16.5 | ND | -0.7 | medium |
| Reed 1987 [47] | 6 | 34 ± 5.3 | 2 | 2 | 36.4 | 6.8 | ND | ND | medium |
| Hill 1979 [49] | 15 | 18 - 45 (range, n = 4) ND (n = 11) | 2 | 2 | 40 | 25 | ND | ND | medium |

[a] In total, 6 studies were included with 1 study split into 2 samples (Hill 1980 NA; Hill 1980 SA), yielding 7 total samples.
[b] Risk of bias (low, medium, high).
All results presented in mean ± SEM.
ND, no data; NA, North American; SA, South African; HF, high-fat; LF, low-fat; TEI, total energy intake.

## 3. Results

### 3.1. Characteristics of included studies

The studies had a total of 206 participants, with a mean age of 46 years, and intervention diets ranged from 2 - 10 weeks (Table 1). The weighted mean difference in fat intake for LF vs HF diets was 20.1% of TEI (LF = 19.5, HF = 39.6), or 580kcal (LF = 547, HF = 1127). 3 studies directly measured and reported TEI [45,46,48], the weighted mean difference for LF vs HF diets was -49 kcal/day (LF = 2877, HF = 2926). 4 studies reported bodyweight [44–46,48], the weighted mean change in bodyweight during the dietary interventions was -0.8kg.

### 3.2. Meta-analyses

All results are presented in standardised mean differences with 95% CIs.

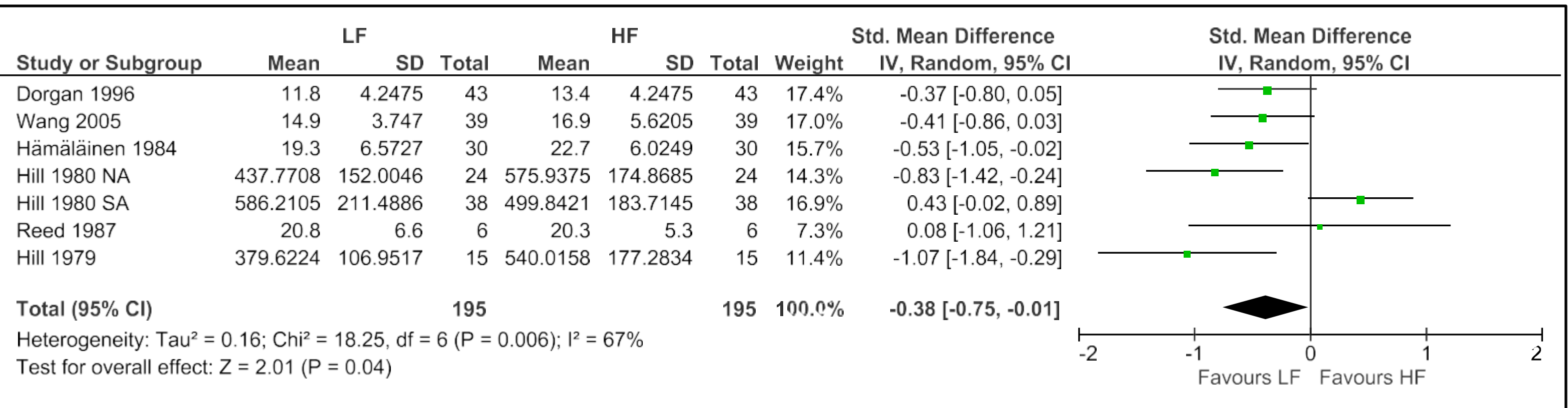

| Study or Subgroup | LF Mean | LF SD | LF Total | HF Mean | HF SD | HF Total | Weight | Std. Mean Difference IV, Random, 95% CI |
|---|---|---|---|---|---|---|---|---|
| Dorgan 1996 | 11.8 | 4.2475 | 43 | 13.4 | 4.2475 | 43 | 17.4% | -0.37 [-0.80, 0.05] |
| Wang 2005 | 14.9 | 3.747 | 39 | 16.9 | 5.6205 | 39 | 17.0% | -0.41 [-0.86, 0.03] |
| Hämäläinen 1984 | 19.3 | 6.5727 | 30 | 22.7 | 6.0249 | 30 | 15.7% | -0.53 [-1.05, -0.02] |
| Hill 1980 NA | 437.7708 | 152.0046 | 24 | 575.9375 | 174.8685 | 24 | 14.3% | -0.83 [-1.42, -0.24] |
| Hill 1980 SA | 586.2105 | 211.4886 | 38 | 499.8421 | 183.7145 | 38 | 16.9% | 0.43 [-0.02, 0.89] |
| Reed 1987 | 20.8 | 6.6 | 6 | 20.3 | 5.3 | 6 | 7.3% | 0.08 [-1.06, 1.21] |
| Hill 1979 | 379.6224 | 106.9517 | 15 | 540.0158 | 177.2834 | 15 | 11.4% | -1.07 [-1.84, -0.29] |
| **Total (95% CI)** | | | **195** | | | **195** | **100.0%** | **-0.38 [-0.75, -0.01]** |

Heterogeneity: Tau² = 0.16; Chi² = 18.25, df = 6 (P = 0.006); I² = 67%
Test for overall effect: Z = 2.01 (P = 0.04)

**Fig. 2.** Total testosterone forest plot, showing the standardised mean difference of low-fat (LF) vs high-fat (HF) diets.

---

There was a small to moderate decrease in TT on LF vs HF diets, which was just statistically significant [-0.38 (95% CI -0.75 to -0.01), P = 0.04] (Fig. 2). There was considerable heterogeneity in the results ($I^2$ = 67%, $Chi^2$ = 18.25, P = 0.006).

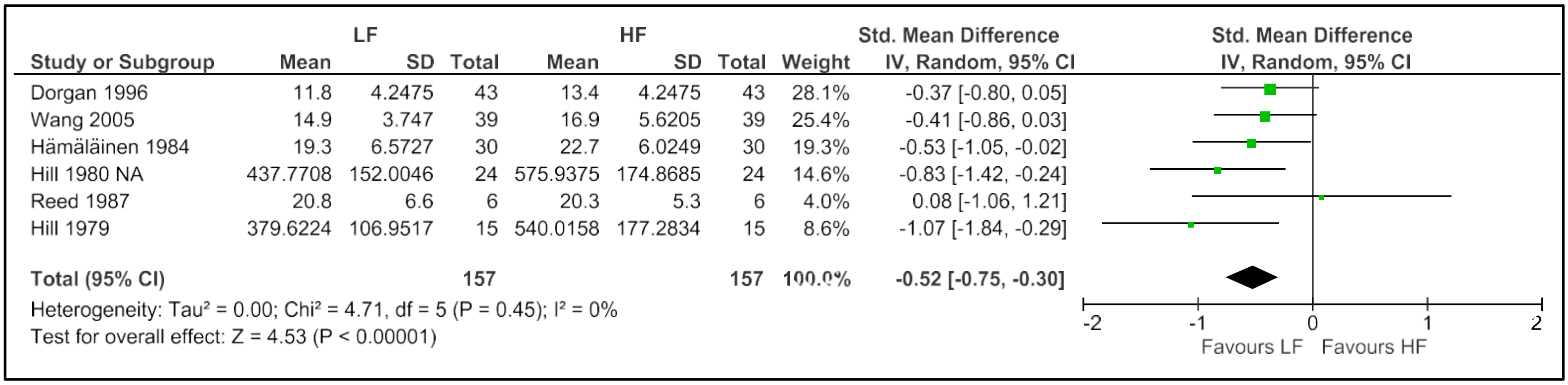

| Study or Subgroup | LF Mean | LF SD | LF Total | HF Mean | HF SD | HF Total | Weight | Std. Mean Difference IV, Random, 95% CI |
|---|---|---|---|---|---|---|---|---|
| Dorgan 1996 | 11.8 | 4.2475 | 43 | 13.4 | 4.2475 | 43 | 28.1% | -0.37 [-0.80, 0.05] |
| Wang 2005 | 14.9 | 3.747 | 39 | 16.9 | 5.6205 | 39 | 25.4% | -0.41 [-0.86, 0.03] |
| Hämäläinen 1984 | 19.3 | 6.5727 | 30 | 22.7 | 6.0249 | 30 | 19.3% | -0.53 [-1.05, -0.02] |
| Hill 1980 NA | 437.7708 | 152.0046 | 24 | 575.9375 | 174.8685 | 24 | 14.6% | -0.83 [-1.42, -0.24] |
| Reed 1987 | 20.8 | 6.6 | 6 | 20.3 | 5.3 | 6 | 4.0% | 0.08 [-1.06, 1.21] |
| Hill 1979 | 379.6224 | 106.9517 | 15 | 540.0158 | 177.2834 | 15 | 8.6% | -1.07 [-1.84, -0.29] |
| **Total (95% CI)** | | | **157** | | | **157** | **100.0%** | **-0.52 [-0.75, -0.30]** |

Heterogeneity: Tau² = 0.00; Chi² = 4.71, df = 5 (P = 0.45); I² = 0%
Test for overall effect: Z = 4.53 (P < 0.00001)

**Fig. 3.** Total testosterone subgroup analysis forest plot (western samples only), showing the standardised mean difference of low-fat (LF) vs high-fat (HF) diets.

---

Compared to TT all samples, the subgroup TT western samples showed a larger decrease in TT on LF vs HF diets [-0.52 (95% CI -0.75 to -0.30) P < 0.001] (Fig. 3). There was also reduced heterogeneity ($I^2$ = 0%, $Chi^2$ = 4.71, P = 0.45), due to the exclusion of Hill 1980 SA, the only sample to show a significant increase in TT on a LF diet [44].

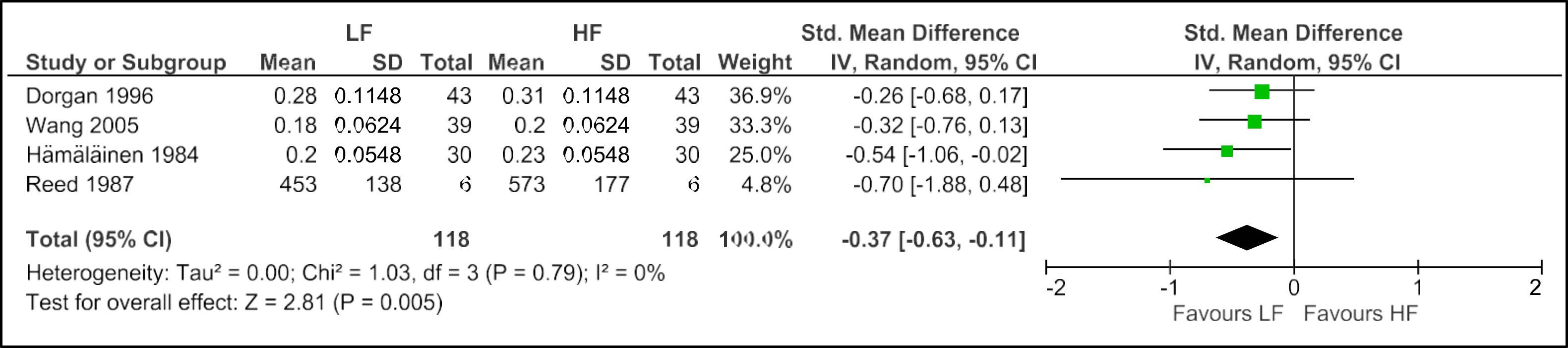

| Study or Subgroup | LF Mean | LF SD | LF Total | HF Mean | HF SD | HF Total | Weight | Std. Mean Difference IV, Random, 95% CI |
|---|---|---|---|---|---|---|---|---|
| Dorgan 1996 | 0.28 | 0.1148 | 43 | 0.31 | 0.1148 | 43 | 36.9% | -0.26 [-0.68, 0.17] |
| Wang 2005 | 0.18 | 0.0624 | 39 | 0.2 | 0.0624 | 39 | 33.3% | -0.32 [-0.76, 0.13] |
| Hämäläinen 1984 | 0.2 | 0.0548 | 30 | 0.23 | 0.0548 | 30 | 25.0% | -0.54 [-1.06, -0.02] |
| Reed 1987 | 453 | 138 | 6 | 573 | 177 | 6 | 4.8% | -0.70 [-1.88, 0.48] |
| **Total (95% CI)** | | | **118** | | | **118** | **100.0%** | **-0.37 [-0.63, -0.11]** |

Heterogeneity: Tau² = 0.00; Chi² = 1.03, df = 3 (P = 0.79); I² = 0%
Test for overall effect: Z = 2.81 (P = 0.005)

**Fig. 4.** Free testosterone forest plot, showing the standardised mean difference of low-fat (LF) vs high-fat (HF) diets.

---

There was a small to moderate, statistically significant decrease in FT on LF vs HF diets [-0.37 (95% CI -0.63 to -0.11) P = 0.005] (Fig. 4). Statistical heterogeneity was low ($I^2$ = 0%, $Chi^2$ = 1.03, P = 0.79). Visually, there was slight heterogeneity in the results, although the direction of effects was consistent.

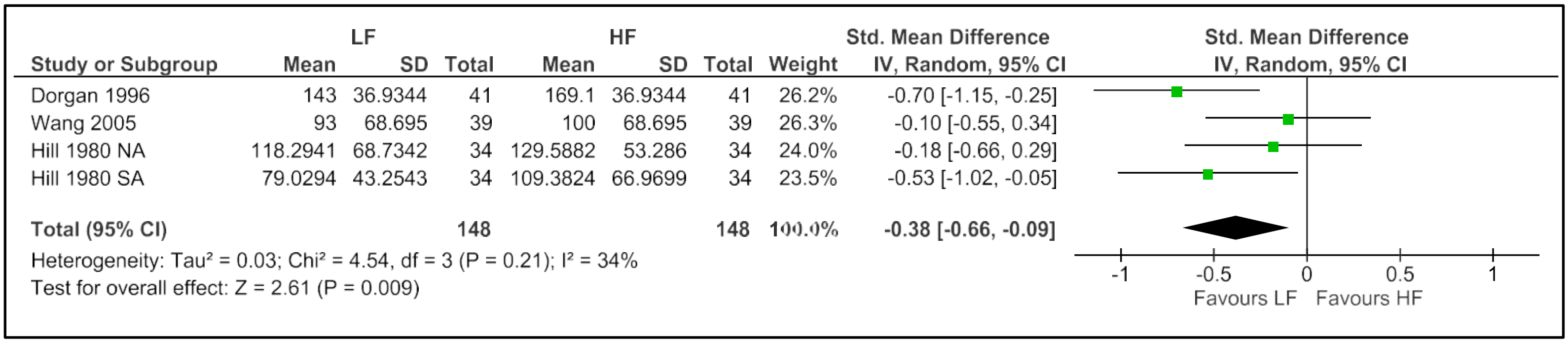

| Study or Subgroup | LF Mean | LF SD | LF Total | HF Mean | HF SD | HF Total | Weight | Std. Mean Difference IV, Random, 95% CI |
|---|---|---|---|---|---|---|---|---|
| Dorgan 1996 | 143 | 36.9344 | 41 | 169.1 | 36.9344 | 41 | 26.2% | -0.70 [-1.15, -0.25] |
| Wang 2005 | 93 | 68.695 | 39 | 100 | 68.695 | 39 | 26.3% | -0.10 [-0.55, 0.34] |
| Hill 1980 NA | 118.2941 | 68.7342 | 34 | 129.5882 | 53.286 | 34 | 24.0% | -0.18 [-0.66, 0.29] |
| Hill 1980 SA | 79.0294 | 43.2543 | 34 | 109.3824 | 66.9699 | 34 | 23.5% | -0.53 [-1.02, -0.05] |
| **Total (95% CI)** | | | **148** | | | **148** | **100.0%** | **-0.38 [-0.66, -0.09]** |

Heterogeneity: Tau² = 0.03; Chi² = 4.54, df = 3 (P = 0.21); I² = 34%
Test for overall effect: Z = 2.61 (P = 0.009)

**Fig. 5.** Urinary testosterone forest plot, showing the standardised mean difference of low-fat (LF) vs high-fat (HF) diets.

---

There was a small to moderate, significant decrease in UT on LF vs HF diets [-0.38 (CI 95% -0.66 to -0.09) P = 0.009] (Fig. 5). There was slight statistical and visual heterogeneity in the results ($I^2$ = 34%, $Chi^2$ = 4.54, P = 0.21), although the direction of effects was consistent.

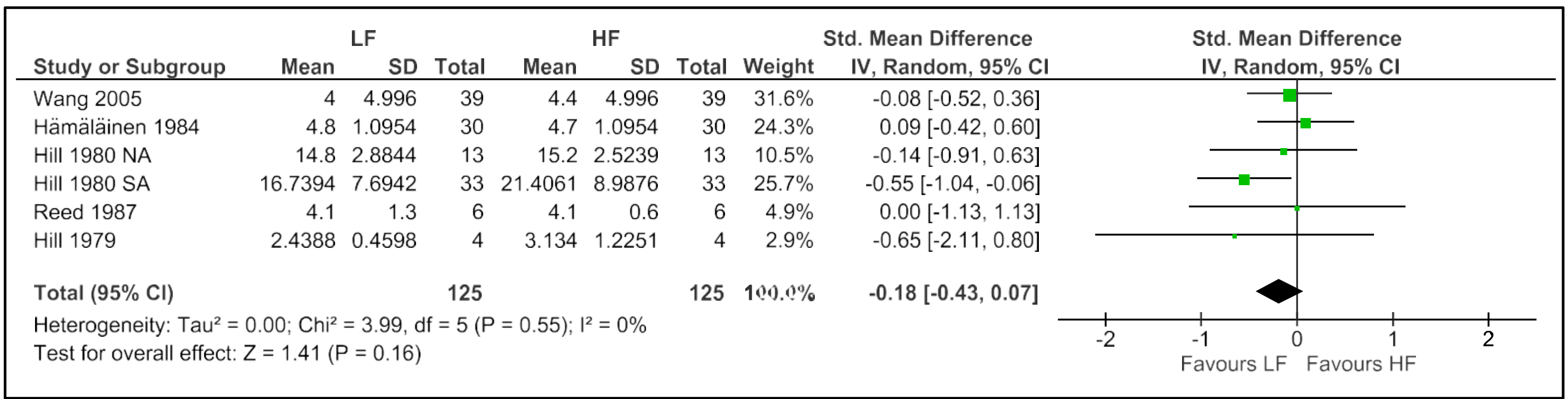

| Study or Subgroup | LF Mean | LF SD | LF Total | HF Mean | HF SD | HF Total | Weight | Std. Mean Difference IV, Random, 95% CI |
|---|---|---|---|---|---|---|---|---|
| Wang 2005 | 4 | 4.996 | 39 | 4.4 | 4.996 | 39 | 31.6% | -0.08 [-0.52, 0.36] |
| Hämäläinen 1984 | 4.8 | 1.0954 | 30 | 4.7 | 1.0954 | 30 | 24.3% | 0.09 [-0.42, 0.60] |
| Hill 1980 NA | 14.8 | 2.8844 | 13 | 15.2 | 2.5239 | 13 | 10.5% | -0.14 [-0.91, 0.63] |
| Hill 1980 SA | 16.7394 | 7.6942 | 33 | 21.4061 | 8.9876 | 33 | 25.7% | -0.55 [-1.04, -0.06] |
| Reed 1987 | 4.1 | 1.3 | 6 | 4.1 | 0.6 | 6 | 4.9% | 0.00 [-1.13, 1.13] |
| Hill 1979 | 2.4388 | 0.4598 | 4 | 3.134 | 1.2251 | 4 | 2.9% | -0.65 [-2.11, 0.80] |
| **Total (95% CI)** | | | **125** | | | **125** | **100.0%** | **-0.18 [-0.43, 0.07]** |

Heterogeneity: Tau² = 0.00; Chi² = 3.99, df = 5 (P = 0.55); I² = 0%
Test for overall effect: Z = 1.41 (P = 0.16)

**Fig. 6.** Luteinising hormone forest plot, showing the standardised mean difference of low-fat (LF) vs high-fat (HF) diets.

---

There was a small non-significant decrease in LH on LF vs HF diets [-0.18 (CI 95% -0.43 to 0.07) P = 0.16] (Fig. 6). Statistical heterogeneity was low ($I^2$ = 0%, $Chi^2$ = 3.99, P = 0.55), however a visual survey of the results revealed moderate heterogeneity and a somewhat inconsistent direction of effects.

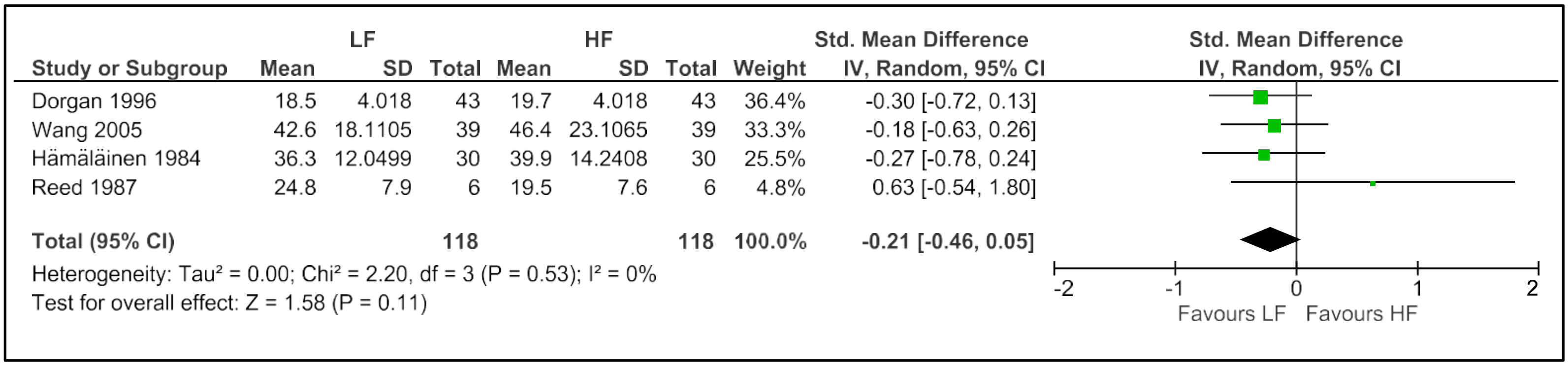

| Study or Subgroup | LF Mean | LF SD | LF Total | HF Mean | HF SD | HF Total | Weight | Std. Mean Difference IV, Random, 95% CI |
|---|---|---|---|---|---|---|---|---|
| Dorgan 1996 | 18.5 | 4.018 | 43 | 19.7 | 4.018 | 43 | 36.4% | -0.30 [-0.72, 0.13] |
| Wang 2005 | 42.6 | 18.1105 | 39 | 46.4 | 23.1065 | 39 | 33.3% | -0.18 [-0.63, 0.26] |
| Hämäläinen 1984 | 36.3 | 12.0499 | 30 | 39.9 | 14.2408 | 30 | 25.5% | -0.27 [-0.78, 0.24] |
| Reed 1987 | 24.8 | 7.9 | 6 | 19.5 | 7.6 | 6 | 4.8% | 0.63 [-0.54, 1.80] |
| **Total (95% CI)** | | | **118** | | | **118** | **100.0%** | **-0.21 [-0.46, 0.05]** |

**Fig. 7.** Sex hormone binding globulin forest plot, showing the standardised mean difference of low-fat (LF) vs high-fat (HF) diets.

---

There was a small non-significant decrease in SHBG on LF vs HF diets [-0.21 (CI 95% -0.46 to 0.05) P = 0.11] (Fig. 7). Both the statistical tests and a visual survey of the results indicated heterogeneity was low ($I^2$ = 0%, $Chi^2$ = 2.20, P = 0.53). Reed 1987 showed qualitive interaction [47], but was an imprecise study, meaning it had low weight in the meta-analysis.

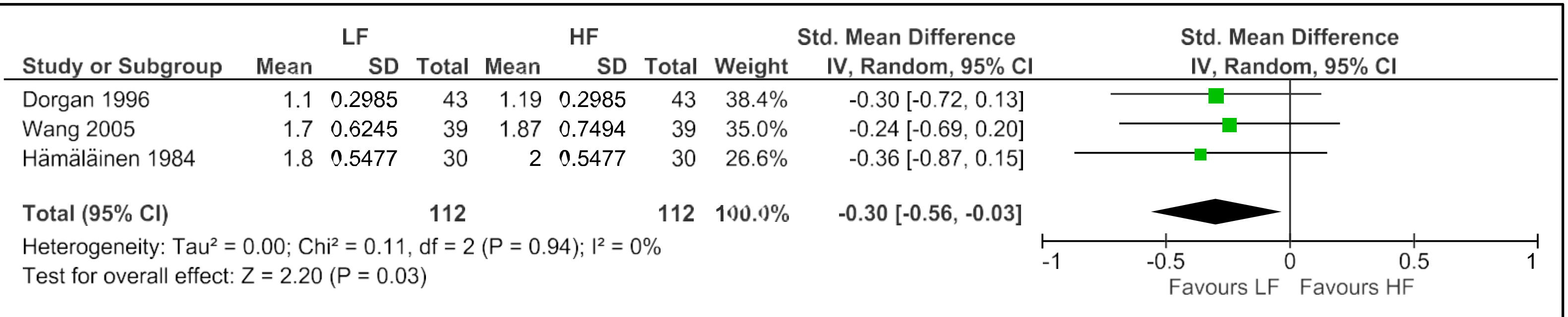

| Study or Subgroup | LF Mean | LF SD | LF Total | HF Mean | HF SD | HF Total | Weight | Std. Mean Difference IV, Random, 95% CI |
|---|---|---|---|---|---|---|---|---|
| Dorgan 1996 | 1.1 | 0.2985 | 43 | 1.19 | 0.2985 | 43 | 38.4% | -0.30 [-0.72, 0.13] |
| Wang 2005 | 1.7 | 0.6245 | 39 | 1.87 | 0.7494 | 39 | 35.0% | -0.24 [-0.69, 0.20] |
| Hämäläinen 1984 | 1.8 | 0.5477 | 30 | 2 | 0.5477 | 30 | 26.6% | -0.36 [-0.87, 0.15] |
| **Total (95% CI)** | | | **112** | | | **112** | **100.0%** | **-0.30 [-0.56, -0.03]** |

**Fig. 8.** Dihydrotestosterone forest plot, showing the standardised mean difference of low-fat (LF) vs high-fat (HF) diets.

---

There was a small significant decrease in DHT on LF vs HF diets [-0.3 (CI 95% -0.56 to -0.03) P = 0.03] (Fig. 8). Both the statistical tests and a visual survey of the results indicated that heterogeneity was low ($I^2$ = 0%, $Chi^2$ = 0.11, P = 0.94).

### 3.3. Sensitivity analyses

**Table 2**
Sensitivity analyses for total testosterone[a]

| Study excluded | Standardised mean difference | 95% CIs | P value from Z test | $I^2$ (%) |
|---|---|---|---|---|
| Dorgan 1996 | -0.39 | -0.85 to 0.08 | 0.1 | 73 |
| Wang 2005 | -0.38 | -0.84 to 0.08 | 0.1 | 72 |
| Hämäläinen 1984 | -0.36 | -0.8 to 0.08 | 0.11 | 72 |
| Hill 1980 NA | -0.31 | -0.7 to 0.09 | 0.13 | 67 |
| Hill 1980 SA[b] | -0.52 | -0.75 to -0.3 | < 0.001 | 0 |
| Reed 1987 | -0.42 | -0.82 to -0.02 | 0.04 | 72 |
| Hill 1979 | -0.29 | -0.67 to 0.08 | 0.12 | 66 |
| Reed 1987 & Hill 1979[c] | -0.33 | -0.73 to 0.08 | 0.12 | 72 |

[a] Sensitivity analyses done by excluding 1 study at a time, and running the meta-analysis again.
[b] This analysis was the same as the subgroup analysis for total testosterone, western.
[c] This analysis excluded the 2 most clinically diverse and highest risk of bias studies.

---

The sensitivity analyses excluding 1 study at a time, all produced similar results except for the analysis minus Hill 1980 SA [44] (Table 2). Excluding this analysis, standardised mean differences were from -0.29 to -0.42, P values were 0.04 - 0.13, and $I^2$ tests were 66 - 73%. This indicates the results for the TT meta-analysis were not unduly reliant on any single study. The sensitivity analyses also confirm the source of the considerable heterogeneity in the TT meta-analysis was Hill 1980 SA [44]. The last sensitivity analysis which excluded the 2 most clinically diverse studies [47,49], produced similar results to all other combinations of studies, besides the analysis minus Hill 1980 SA [44]. This indicates clinical heterogeneity did not have a significant effect on the TT meta-analysis.

### 3.4. Risk of bias

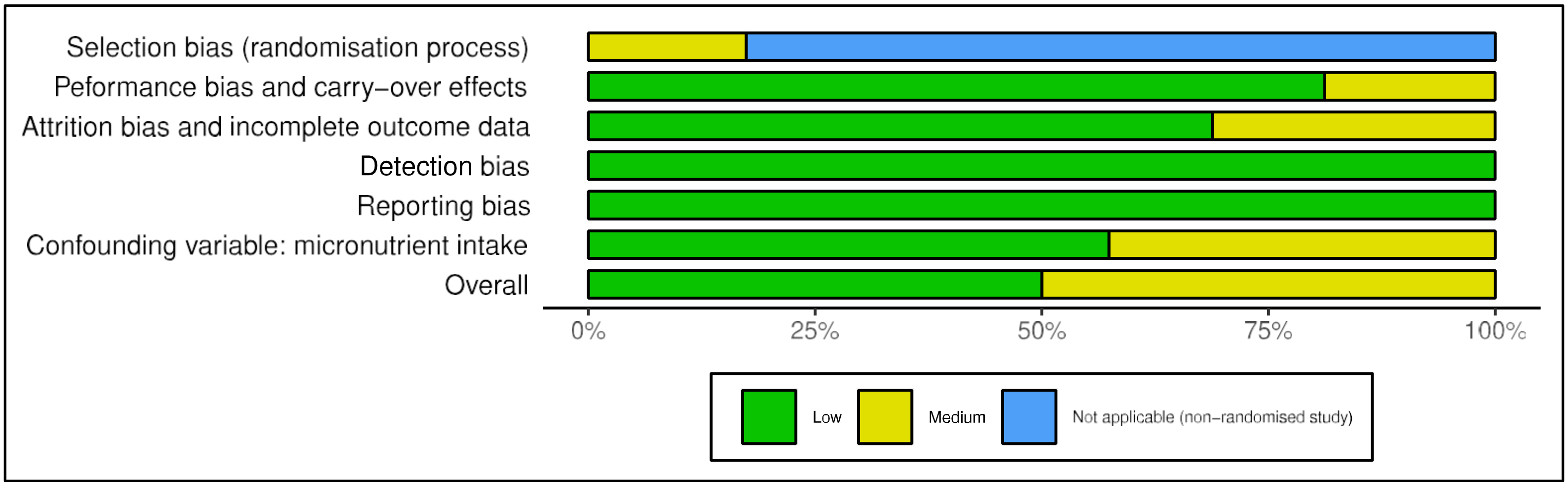

**Fig. 9.** Risk of bias graph, showing risk of bias at the review level, stratified by bias domains.

---

3 studies were at low risk of bias [45,46,48], and 3 studies were at medium risk of bias [44,47,49] (Table 1). In Reed 1987 and Hill 1979, unmeasured changes in calorie intake may have altered the effect of LF diets on outcomes, although these studies stated that the intervention diets were isocaloric [47,49] (Fig. 9 - performance bias and carry-over effects). Hill 1980 NA and SA may have lacked full outcome data for all participants, nevertheless all outcome data reported showed a consistent pattern of effects [44,61–63] (Fig. 9 - attrition bias and incomplete outcome data). The forest plots were ordered from lowest to highest risk of bias, which revealed a visual trend of smaller more consistent effects in low-bias studies, and larger more inconsistent effects in medium-bias studies. This was particularly evident in the forest plots for TT, FT and SHBG (Figs. 2, 4, and 7).

## 4. Discussion

The main findings of this review were that LF vs HF diets resulted in small to moderate decreases in TT, FT, UT, and DHT in men (Figs. 2, 4, 5, and 8). There was a small decrease in bodyweight during the dietary interventions (-0.8kg), which is unlikely to have significantly affected outcomes. Most of the studies used a HF/LF diet order, with the HF period mimicking the participants' baseline diet. The most notable exception to this was Hill 1980 SA which used a LF/HF diet order and was the only study to find a significant increase in TT on a LF diet [44] (Fig. 2). Thus, it could be argued that any dietary change, regardless of macronutrient composition results in a decrease in androgens. However, in Hill 1980 SA UT substantially increased on the HF diet which does not indicate lowered androgen status [62] (Fig. 5). Also, Dorgan 1996 used a 2-group crossover design (AB/BA), and found diet order made no difference to changes in outcomes [48]. Similarly, both Hämäläinen 1984 and Wang 2005 included a 3rd HF switchback diet, and found no effect of diet order on outcomes [45,46].

### 4.1. Mechanisms: main findings

The results for TT showed a high degree of heterogeneity ($I^2$ = 67%) (Fig. 2), which was decreased in the subgroup TT western ($I^2$ = 0%) (Fig. 3). This suggests ethnic and genetic or epigenetic variation in TT, in response to dietary fat intake. The remaining visual heterogeneity in TT western may be attributable to a variety of factors, one of which being differences in micronutrient intake (Fig. 9). The largest decreases in TT were seen in the 2 studies with vegetarian LF diets (Hill 1979; Hill 1980 NA) [44,49]. These diets may have been lower in zinc, which is a common feature of vegetarian diets [64], and marginal zinc deficiency has been found to decrease TT [32]. Nevertheless, studies well matched for micronutrient intake showed similar, albeit smaller changes in TT [45,48] (Fig. 2); suggesting the decrease in TT was mostly due to other dietary factors.

Dietary fibre intake was likely higher on LF vs HF diets, which has been suggested to increase T excretion by modulating the enterohepatic circulation of steroids [25]. However, we found LH (P = 0.16) and UT excretion (P = 0.009) decreased on LF diets, which suggests decreased T production rather than increased T excretion (Figs. 5 and 6). Moreover, using a 12hr trideuterated infusion of T, Wang 2005 found no change in T excretion on the LF diet, but decreased T production [46]. 2 studies measured follicle-stimulating hormone, which showed inconsistent effects on LF diets [46,61]. Estradiol was measured in 4 and estrone in 2 studies, via blood sampling [44–46,48,61]. The results showed either non-significant changes (mostly decreases), or significant decreases on LF diets. This suggests an upregulation of aromatase, leading to increased estrogens was not responsible for the lower T on LF diets. In meta-analysis, DHT significantly decreased on the LF diets, which indicates decreased T production, rather than an inhibition of 5α-reductase leading to a build-up of T (Fig. 8). We found weak evidence of a small decrease of SHBG on LF diets (Fig. 7). This suggests the decrease in FT on LF diets was largely due to lower TT, rather than higher SHBG bound T. To summarise, our findings indicate that endogenous T production decreased on LF diets, leading to lower FT and TT.

The HF diets had increased dietary cholesterol and caused increased blood cholesterol. Since, T is synthesised from cholesterol it is logical to think that increased cholesterol substrate, increased T production. However, in men hypercholesterolemia is associated with lower TT [65]; and in rodents high cholesterol diets decrease TT by downregulating steroidogenic enzymes [66]. Similarly, the HF vs LF diets likely had higher dietary arachidonic acid, due to higher intakes of animal foods. In vitro, exogenous arachidonic acid has been shown to increase T production in Leydig cells [67]; however arachidonic acid supplementation in men has not been found to affect TT or FT [68].

The LF vs HF diets were consistently lower in monounsaturated fatty acids (MUFA) and saturated fatty acids (SFA), and had higher polyunsaturated to saturated fatty acid ratios (P:S). This suggests a beneficial effect of MUFA and SFA, and/or a deleterious effect of polyunsaturated fatty acids (PUFA) on androgens. A similar but ineligible study found that decreasing MUFA and SFA, and increasing P:S whilst keeping total fat intake stable, decreased TT by 15% [69]. The beneficial effect of MUFA intake on T is supported by another study which replaced 25g/day butter with either olive or argan oil, and found TT increased by 17.4% and 19.9% respectively ($P < 0.001$) [70]. In rodents, fatty acid intake strongly modifies testicular lipid composition. High PUFA vs MUFA or SFA diets result in decreased T production via increased testicular oxidative stress, decreased steroidogenic enzymes and decreased testicular free cholesterol available for steroidogenesis [71,72]. For ethical reasons, similar experiments have not been conducted in humans. However, intervention and cross-sectional studies have found that blood and adipose lipids similarly reflect dietary intake, with stronger effects for PUFA [73]. High intakes of linoleic acid, the main dietary omega-6 PUFA, have been shown to increase markers of oxidative stress in men [74]. Oxidative stress is well known to adversely affect semen parameters [75]; and this effect may extend to testicular steroidogenesis. Omega-6 intake has been inversely correlated to testicular volume, suggesting a direct adverse effect on testicular function [22]. Thus, the decrease in MUFA and SFA intake, and relative increase in omega-6 PUFA on LF diets, may have altered testicular lipid composition and increased oxidative stress, thereby decreasing T production.

### 4.2. Mechanisms: subgroups

Androgen metabolism likely differs by ethnicity, as ethnic differences have been found in men's FT, SHBG and follicle-stimulating hormone levels [76,77]. Whether androgen metabolism differs by ethnicity in response to diet has not been well studied. However, the limited results from our subgroup analysis suggests that the decrease in TT on LF diets is larger and more consistent in western vs non-western men (Figs. 2 and 3). Palaeolithic ancestors of modern Europeans likely had a HF intake, as reliance on animal foods in 20th century hunter gatherers increases ≥40° latitude, putting almost all of Europe in this category [78]. In addition, modern Europeans have a high prevalence of the -13,910 C>T allele which gives rise to the phenotype of lactase persistence, and confers the ability to digest lactose throughout adulthood [79]. This suggests that after the Neolithic revolution Europeans continued to enjoy HF intakes, by consuming traditional HF dairy foods [80]. Since genetic adaptations to environmental changes occur slowly, the majority of human genes remain unchanged since the upper Palaeolithic and early Neolithic periods [81]. Thus, men with European ancestry may have other genetic adaptations that promote a survival or reproductive advantage in response to a HF diet. Therefore, we hypothesize that T levels are adversely affected by a LF diet in men of European descent.

### 4.3. Secular decline in testosterone

In the USA, from 1965-91 fat intake decreased by 10.1% of TEI [15], and the first evidence of a secular decline in T follows this [5]. The decline in fat intake largely came from decreases in SFA and MUFA, whilst PUFA intake remained relatively stable [82]. The change in fat intake from 1965-91, is comparable to the mean difference between the HF vs LF diets of this review, although decreases in fat intake were larger in the reviewed studies (10.1 vs 20.1% of TEI). The results of this review suggest that the decline in fat intake over the 1970s and 80s, may have adversely affected men's TT and FT. However, since fat intake had stabilised by 1991 [15], it is unlikely to explain the continual secular decline in T. To our knowledge, it is unknown whether LF diets have a larger or smaller effect in the long term, due to a lack of longitudinal studies measuring fat intake and T.

### 4.4. Limitations

The main limitation of the review was the low number of studies and small total sample size (n = 206). This contributed to the somewhat larger CIs and P values than expected for reasonably homogeneous studies. Additionally, we did not use the variance of within participant differences for the pooled effect estimates; thus it is possible that the effect estimates had slightly larger CIs than true [50]. In contrast, our method for the variance of effects was more conservative [50], and so increased the confidence in our results. It also avoided possibly inaccurate changes in the weighting of studies in the meta-analyses. A separate issue was that the DerSimonian and Laird method [52] arguably performs poorly with a low number of studies [83], which meant our heterogeneity tests were unpowered and CIs possibly too narrow. However, since statistical heterogeneity was low, besides TT, and there was no clear-cut alternative method [84], we chose to use the conventional DerSimonian and Laird approach [52]. To alleviate these problems, we visually inspected forest plots to detect additional heterogeneity. The low number of studies also meant that we were unable to produce a meaningful funnel plot to investigate publication bias.

We attempted to explain the considerable heterogeneity in TT by doing a post hoc subgroup analysis based on ethnicity (Fig. 3). However, the subgroup western included a small number of NA black men (12.7%), meaning it did not uniformly represent men of white/ European descent, although they were the majority (87.3%). Also, it is possible that the study excluded in the subgroup analysis [44], was an outlier for another reason such as some facet of study design or chance. Subgroup analyses are observational in nature, particularly when done post hoc, so this process cannot produce firm conclusions and is best used to produce hypotheses [55]. A strength of the subgroup analysis is that it relied upon within study differences in effects, rather than between study differences. Also, it was based on qualitive interaction rather than a difference in magnitude of effect. The low number of studies made statistical and qualitative explanations of heterogeneity hard to distinguish from chance, particularly in outcomes with less than 7 samples. Additional sources of heterogeneity including differences in micronutrient and fatty acid intake could not be robustly explored, due to a lack of study data and the low number of studies.

Another limitation was that half of the studies included were at medium risk of bias (Table 1 and Fig. 9). In addition, only 1 study was a fully randomised (AB/BA) crossover trial [48], which increased the risk of selection and performance bias, particularly at the review level (Fig. 9). Across outcomes, the low vs medium-bias studies showed smaller more consistent effects. This was most evident in DHT which included only the 3 low-bias studies [45,46,48], and showed very little heterogeneity visually or statistically ($I^2$ = 0%; $Chi^2$ = 0.11, P = 0.94) (Fig. 8). A strength of the review was that the strict eligibility criteria provided studies relatively free from confounding variables, and the extensive search strategy likely captured all eligible studies.

Throughout the review, the primary measurement of fat intake used was relative intake, measured as a percentage of TEI. Since only weight-maintaining isocaloric dietary interventions were included, percentage fat intake provided a measure of absolute fat intake, which accounted for participant differences in energy intake requirements. However, this approach meant we were unable to detect the effects of absolute fat intake on outcomes, irrespective of participant energy requirements. An alternative approach would have been to conduct a correlation-based analysis using absolute fat intake vs outcomes for each individual diet (or participant). However, due to the differences in assays and plasma vs serum measurements, this was not appropriate for the included studies. Also, this analysis would have ignored the relative within-study effects on outcomes. A full discussion of the merits of using absolute, rather than relative fat intake is beyond the scope of this review.

## 5. Conclusions

The results of this review suggest that LF vs HF diets moderately decrease T levels in men, via a reduction in testicular T production. However, due to the small total sample size (n = 206), heterogeneity in TT (Fig. 2), and risk of bias (Fig. 9); large randomised controlled trials are needed to confirm this review’s findings, before practical recommendations can be made. Ideally, such studies would also investigate the effects of ethnicity on androgen metabolism, in response to dietary fat intake.

**Funding**

This research did not receive any specific grant from funding agencies in the public, commercial, or not-for-profit sectors.

**CRediT authorship contribution statement**

**Joseph Whittaker**: Conceptualization, Methodology, Validation, Formal analysis, Investigation, Resources, Data Curation, Writing - Original Draft, Writing - Review & Editing, Visualization, Project administration. **Kexin Wu:** Methodology, Formal analysis, Resources, Data Curation, Writing - Review & Editing, Visualization.

**Declaration of Competing Interest**

The authors report no declarations of interest.

**Acknowledgements**

The authors would like to thank Miranda Harris, Justine Bold and Lindsey Fellows from the University of Worcester, for their feedback on drafts; and Keith Whittaker for proof reading. This work was partly based on an unpublished Master's thesis, completed at the University of Worcester.